# The Quantum Confinement Effect on the Spectrum of Near-Field Thermal Radiation by Quantum Dots

Saman Zare and Sheila Edalatpour

*Department of Mechanical Engineering, University of Maine, Orono, Maine 04469, USA*

*Frontier Institute for Research in Sensor Technologies, University of Maine, Orono, Maine 04469, USA*

**ABSTRACT**. The quantum confinement effect on the spectrum of near-field thermal radiation by periodic and random arrays of quantum dots (QDs) is investigated. The local density of states (LDOS) thermally emitted by QD arrays made of three lead chalcogenides, namely, lead sulfide, lead selenide, and lead telluride, is computed at a near-field distance from the arrays. The dielectric function of the QDs is extracted from their absorption spectra by utilizing an optimization technique. The thermal discrete dipole approximation is used for computing the LDOS. It is shown that the peak wavenumber of near-field LDOS emitted by periodic arrays of lead chalcogenide QDs can be significantly modulated (up to 4490 cm$^{-1}$) by varying the size of the dots. The LDOS is proportional to the imaginary part of the QDs' polarizability which peaks at the bandgap energy of the QDs. The bandgap energy of the QDs (and thus the LDOS peak) is significantly affected by the quantum confinement effect which is size-dependent. While the magnitude of thermal radiation by random arrays of QDs can be different from the periodic arrays with the same filling factor by up to ±26%, the LDOS spectrum and peak location are the same for both periodic and random arrays. The peak wavenumber of near-field radiative heat transfer between the QD arrays is also strongly affected by quantum confinement in the QDs, and thus it can be tuned by changing the size of the QDs.

# I. INTRODUCTION

Thermal radiation at sub-wavelength distances from an emitting medium, which is referred to as the near-field thermal radiation, can exceed the far-field blackbody limit by several orders of magnitude and be quasi-monochromatic[1]. Due to these unique properties, near-field thermal radiation has attracted significant attention for many promising applications in imaging[2-7], energy conversion using near-field thermophotovoltaic (TPV) devices[8-24], nanomanufacturing[25-30], thermal management of electronic devices[31-33], and thermal rectification[34-46]. Near-field applications such as near-field TPV power generation and thermal rectification require tuning the spectrum of near-field thermal radiation. One mechanism that has been widely explored for customizing the near-field spectra is engineering the emitting medium at the sub-wavelength scale. Thermal radiation of these engineered materials, known as metamaterials, significantly differ from those of the corresponding bulk materials and can be tuned by modifying the geometry and size of the sub-wavelength features. A variety of metamaterials, such as periodic gratings[47-67], photonic crystals[68-74], hyperbolic metamaterials[75-104], nanoporous metamaterials[53,57,105-108], and double-negative metamaterials[101,109-133], have been proposed for tuning the near-field spectra. The spectrum of near-field thermal radiation by gratings can be tuned by varying the material properties and the geometry of the grating pattern[47-67]. Depending on the geometry and material properties, various modes such as transverse magnetic (TM) and transverse electric (TE) guided modes[52,55], spoof surface polaritons[52,55], hyperbolic modes[54], and magnetic polaritons[56] can be excited in gratings, resulting in near-field radiative properties different from those of the bulk material. Photonic crystals[68-74] are sub-wavelength periodic structures engineered to have a band gap that forbids propagation of a certain frequency range of thermal radiation. This property has been capitalized on for frequency-selective near-field thermal emission. Different mechanisms such as increased contribution from frustrated total internal reflection modes[73], band-folding effect[72], surface Bloch waves coupling[69], and the resonant leaky modes[70] can contribute to the near-field thermal emission depending on the geometry and material properties of the photonic crystal. Hyperbolic metamaterials[75-104] are anisotropic materials with metallic behavior (negative permittivity) in one direction and dielectric behavior (positive permittivity) in other directions. Thermal radiation by hyperbolic metamaterials shows a broadband enhancement due to the contribution of electromagnetic waves with very large wavevectors. Nanoporous materials[53,57,105-108] are anisotropic materials which can support a surface mode in addition to the one supported by the isotropic bulk material. The spectral location of the surface modes can be tuned by varying the filling fraction of the pores. Double-negative metamaterials[101,109-133] have sub-wavelength inhomogeneities that are arranged in repeating patterns with a sub-wavelength periodicity. These metamaterials support both TM- and TE-polarized surface modes. The TE-polarized surface modes, which are not supported by most natural materials, cause an additional peak in the spectrum of near-field thermal radiation. Fabricating thermal metamaterials usually requires advanced, costly and time-consuming manufacturing techniques.

Another less explored but promising mechanism for tuning the spectrum of near-field thermal radiation is modulating the electronic structure of the emitter by capitalizing on the quantum confinement effect. The quantum confinement effect arises when the size of the emitter becomes comparable to the atomic length scales at least in one dimension. Quantum dots (QDs) are an example of these quantum-sized emitters which experience quantum confinement in three dimensions. They are quasi-spherical nanocrystals composed of tens to a few thousand atoms[134]. Because their dimensions are comparable to or smaller than the Bohr exciton radius, the electronic structure of QDs becomes quantized in the form of an ensemble of discrete energy levels. This quantization of energy levels results in modification of the bandgap and electromagnetic properties of the QDs compared to the bulk material in which energy levels are continuous[134]. The level of the quantum confinement depends on the size of the QDs[135]. As the size of the QDs decreases, the energy levels become more distant, and the bandgap increases. As such, the bandgap and electromagnetic properties of the QDs can be modulated by changing the size of the quantum dots. The size-dependent electromagnetic properties of quantum materials present an opportunity for designing thermal emitters with desired near-field emission spectra. In addition, QDs can be manufactured in bulk quantities and with high quality and size monodispersity using cost-efficient, solution-based techniques[136]. The opportunity for tuning the spectrum of near-field thermal radiation by capitalizing on the quantum confinement effect has not been explored. In this paper, we compute the near-field thermal emission spectra of QD arrays and show that the spectral location of the peak of thermal radiation can be tuned over a wide wavenumber range by changing the size of the QDs.

## II. PROBLEM DEFINITION AND METHODS

A schematic of the problem under consideration is shown in Fig. 1. A periodic array of QDs with diameter $D$ and temperature $T$ is emitting into the free space. The QDs in the array are located on a rectangular grid with a pitch size of $L$. The LDOS thermally emitted at an observation distance $d$ above the array and along the central axis of the QDs is to be calculated. Arrays of three lead chalcogenides QDs, namely lead sulfide (PbS), lead selenide (PbSe), and lead telluride (PbTe) QDs, are selected for this study. The Bohr exciton radii of lead chalcogenides are large (~18 nm for PbS[137], ~46 nm for PbSe[138], and ~150 nm for PbTe[139]), and therefore strong quantum confinement can be observed even for relatively large QDs of these materials. This allows tuning the spectral location of the peak of thermal emission over a wide range of wavenumbers. The bandgap of lead chalcogenides QDs is located in the wavenumber range of 5000-10000 cm$^{-1}$ [140-144], which can be thermally excited. Lead chalcogenides QDs are more suitable for high-temperature emitters. For low temperate applications, mercury chalcogenides QDs[145] can be considered.

The electromagnetic (and thus thermal radiative) properties of QDs are determined from their dielectric function, $\varepsilon$, which is dependent on the size of the QDs. The size-dependent dielectric function of the QDs can be extracted from their measured absorbance spectra[140,141,143] by utilizing the Maxwell-Garnett effective medium theory (EMT) and the Kramers-Kronig (KK) relation[146]. The absorption spectra of dilute solutions (filling fractions of ~$10^{-3}$ – $10^{-5}$) of lead chalcogenides quantum dots in a transparent solvent have been measured using spectroscopy techniques[140,141,143]. The solvent for PbS and PbTe QDs is tetrachloroethylene ($C_2Cl_4$), while carbon tetrachloride ($CCl_4$) has been used for the PbSe QDs. For dilute solutions of QDs, the volume fraction of the QDs is very low and thus the absorption coefficient, $\alpha_\lambda$, of the solution can be related to the dielectric function of the QDs using the Maxwell-Garnett EMT as[147]:

$$\alpha_\lambda = \frac{18\pi}{\lambda} \frac{n_s^3}{\left(\varepsilon' + 2n_s^2\right)^2 + \varepsilon''^2} \varepsilon'' \tag{1}$$

where $\lambda$ is the wavelength, $n_s$ is the solvent refractive index (1.53 for $C_2Cl_4$[148] and 1.46 for $CCl_4$[148]), and $\varepsilon'$ and $\varepsilon''$ are the real and imaginary parts of the dielectric function of the QDs, respectively. The measured absorbance spectra have arbitrary units, and therefore do not directly provide the magnitude of the absorption coefficient, $\alpha_\lambda$. The magnitude of $\alpha_\lambda$ is determined from the measured absorption spectra using the following procedure. It has been experimentally observed that the quantum confinement effect in the lead chalcogenide QDs is negligible at wavelengths below 400 nm[149]. As such, the dielectric function of the lead chalcogenide QDs for wavelengths shorter than 400 nm is equal to that of the bulk. This allows calculating $\alpha_\lambda$ of the QD solutions in the spectral range of 0 – 400 nm using the dielectric function of the bulk and Eq. (1) [140,141]. For wavelengths longer than 400 nm, first the absorbance spectra are normalized by their value at $\lambda = 400$ nm. Then, the normalized spectra are multiplied by the absorption coefficient at 400 nm, $\alpha_{\lambda=400nm}$, as found using Eq. (1) and the dielectric function of the bulk, to obtain $\alpha_\lambda$ for $\lambda > 400$ nm.

Equation (1) provides a relation between $\alpha_\lambda$ and the dielectric function of the QDs, $\varepsilon$. However, the knowledge of $\alpha_\lambda$ is not sufficient for finding $\varepsilon$ as there are two unknowns, namely $\varepsilon'$ and $\varepsilon''$, in Eq. (1). A relation between $\varepsilon'$ and $\varepsilon''$ can be established via the KK relation. The KK relation provides an equation for determining $\varepsilon'$ at a given wavelength when $\varepsilon''$ is known for all other wavelengths. To use the KK relation, first a wavelength interval between 0 and an upper limit, $\lambda_{max}$, is selected. The upper limit is selected such that $\alpha_\lambda$ is negligible for $\lambda > \lambda_{max}$. When $\alpha_\lambda$ is very small, $\varepsilon''$ is negligible and it does not contribute to $\varepsilon'$. The wavelength interval between 0 and $\lambda_{max}$ is then discretized into $N_\lambda$ equal subintervals of length $\Delta\lambda$. The wavelength subintervals are small enough such that the peaks and dips in the absorbance coefficient are captured in the discretized data. Finally, the discretized KK relation is used to relate the real part of the dielectric function at a given wavelength $\lambda_j$ (i.e., $\varepsilon'(\lambda_j)$) to the imaginary part of the dielectric function at all other wavelengths (i.e, $\varepsilon''(\lambda_k)$, where $k = 1, 2, \ldots, N_\lambda+1$ and $k \neq j$) as [146]:

$$\varepsilon'(\lambda_j) = \varepsilon_\infty + \frac{2}{\pi} \sum_{\substack{k=1 \\ k \neq j}}^{N_\lambda+1} \frac{\lambda_j^2 \Delta\lambda}{\lambda_k(\lambda_j^2 - \lambda_k^2)} \varepsilon''(\lambda_k) \qquad (2)$$

where $N_\lambda+1$ is the number of discretized wavelengths, and $\varepsilon_\infty$ is the high-frequency dielectric function which equals 1.5 for PbSe[150] and 1.7 for PbS[151] and PbTe[152].

The real and imaginary parts of the dielectric function of the QDs are found using Eqs. (1) and (2) and by utilizing an optimization technique. A block diagram of the optimization technique is shown in Fig. 2. The optimization process starts with an initial guess for the imaginary part of the dielectric function of QDs, $\varepsilon''$, at all discretized wavelengths. Then, the real part of the dielectric function, $\varepsilon'$, at a given wavelength $\lambda_j$ ($j$ = 1, 2, …, $N_\lambda$+1), is computed from the imaginary part at all other wavelengths, $\lambda_k$ ($k$ = 1, 2, …, $N_\lambda$+1, $k \neq j$), using the KK relation (Eq. (2)). Using the initial guess for $\varepsilon''$ and the computed $\varepsilon'$, the absorption coefficients at the discretized wavelengths, $\alpha_{\lambda_j}$ is estimated by Eq. (1). The total error involved in estimating $\alpha_\lambda$ is defined as the summation of the squared difference of the measured ($\alpha_{\lambda_j,meas}$) and computed ($\alpha_{\lambda_j,comp}$) absorption coefficients over the discretized wavelengths, i.e.,

$$err = \sum_{j=1}^{N_\lambda+1} \left( \alpha_{\lambda_j,comp} - \alpha_{\lambda_j,meas} \right)^2 \qquad (3)$$

Based on the error found from Eq. (3), $\varepsilon''$ is corrected for all discretized wavelengths. Then, $\varepsilon'$ and $\alpha_\lambda$ are reevaluated in another iteration until the difference between $\varepsilon''$ in two consecutive iterations becomes less than $10^{-6}$ for all discretized wavelengths. To implement this optimization process, MATLAB's built-in *fmincon* function, which is a constrained nonlinear and multivariable minimization function, is utilized in which the interior-point method (IPM) is used for correcting $\varepsilon''$ in each iteration. The objective function of *fmincon* is the total error defined by Eq. (3), while the decision variables are the imaginary parts of the dielectric function $\varepsilon''$ at the discretized wavelengths.

Once the size-dependent dielectric function of the QDs is found, the local density of states (LDOS) thermally emitted by a periodic array of QDs at a near-field observation distance from the array is computed using the periodic thermal discrete dipole approximation (T-DDA)[153]. In the periodic T-DDA, the unit cell of the array (one quantum dot in this study) is discretized into cubical sub-volumes with sizes much smaller than the wavelength, the observation distance, and the interdot spacing. In this case, the electric field within each sub-volume can be assumed as uniform (i.e., the sub-volumes behave as electric point dipoles)[153]. Then, the electric (magnetic) Green's function, $\mathbf{g}_{i,o}^{E(H)}$, representing the electric (magnetic) field generated in the unit-cell sub-volumes due to a point source at the observation location is obtained by solving a discretized form of Maxwell's equations as[153]:

$$\frac{1}{\alpha_i} V_i \varepsilon_0 (\varepsilon_i - 1) \mathbf{g}_{i,o}^{\gamma} - k_0^2 \sum_{j=1}^{N} V_j (\varepsilon_j - 1) \mathbf{G}_{i,j}^{0E,P} \cdot \mathbf{g}_{j,o}^{\gamma} = \mathbf{G}_{i,o}^{0\gamma,P} , \quad \gamma = E \text{ or } H, \ i = 1, 2, ..., N \qquad (4)$$

where $\alpha_i$, $V_i$ and $\varepsilon_i$ are the polarizability, volume and dielectric function of sub-volume $i$, respectively, $\varepsilon_0$ is the free-space permittivity, $k_0$ is the magnitude of the wavevector in the free space, subscript $o$ refers to the observation point, and $\mathbf{G}_{k,l}^{0E(H),P}$ is the electric (magnetic) periodic Greens function between points $k$ and $l$ in the free space[153]. It should be noted that the Greens function $\mathbf{g}_{i,o}^{E(H)}$ is wavevector-dependent, and thus Eq. (4) should be solved for all wavevectors in the Brillouin zone (i.e., wavevectors between -π/L and π/L). The Green's functions for the replica $(m, n)$ of sub-volume $i$ in the unit cell can be found using $\mathbf{g}_{i,o}^{E(H)}$ as[153]:

$$\mathbf{g}_{imn,o}^{\gamma} = \mathbf{g}_{i,o}^{\gamma} e^{iL(mk_x + nk_y)}, \ m, n = 0, \pm 1, \pm 2, ... \qquad (5)$$

Then, the total electric (magnetic) Greens function between replica $(m, n)$ of sub-volume $i$ and the observation point $o$, $\mathbf{G}_{imn,o}^{E(H)}$, can be found from the wavevector-dependent electric (magnetic) Greens functions as:

$$\mathbf{G}_{imn,o}^{\gamma} = \frac{L^2}{4\pi^2} \int_{-\pi/L}^{\pi/L} \int_{-\pi/L}^{\pi/L} \mathbf{g}_{imn,o}^{\gamma}(k_x, k_y) dk_y dk_x , \ \gamma = E \text{ or } H \qquad (6)$$

Finally, the spectral LDOS emitted by the array at the observation distance, $\rho_\omega$, is found using the total electric and magnetic Greens functions computed using Eq. (6) as:

$$\rho_\omega = \frac{2k_0^2}{\pi\omega} \sum_{i=1}^{N} V_i \varepsilon_i'' \sum_{m=0}^{N_k} \sum_{n=0}^{N_k} Trace\left[ k_0^2 \mathbf{G}_{imn,o}^E \otimes \mathbf{G}_{imn,o}^E + \mathbf{G}_{imn,o}^H \otimes \mathbf{G}_{imn,o}^H \right] \quad (7)$$

where $\omega$ is the angular frequency, $N$ is the number of sub-volumes used for discretizing the unit cell of the array, $\otimes$ shows the outer product, and $N_k$ is the number of wavevectors used for discretizing the Brillouin zone.

## III. RESULTS AND DISCUSSION

The dielectric function of PbS, PbSe, and PbTe QDs with various diameters $D$ ranging from 3.3 nm to 6.8 nm is extracted from their absorption spectra using the optimization technique described in Section II (see Fig. 2). Several initial guesses, including $\varepsilon''$ of the bulk material ($\varepsilon''_{bulk}$), a constant (over wavelength) $\varepsilon''$ equal to $\varepsilon''_{bulk}$ at $\lambda = 400$ nm, and a $\varepsilon''$ inversely proportional to $\lambda$ as $\varepsilon''(\lambda) = \frac{400\,\text{nm}}{\lambda(\text{in nm})} \varepsilon''_{bulk}(400\,\text{nm})$, are used for the decision variable when running the *fmincon* function. All initial guesses result in the same dielectric function for QDs. However, a faster convergence is achieved when $\varepsilon''$ of the bulk material is used as the initial guess. The $\varepsilon''$ of the bulk materials is obtained from literature[150-152].

As an example, the real and imaginary parts of the dielectric function of PbSe QDs are shown in Figs. 3a and 3b, respectively. The dielectric function of a bulk of PbSe[150] is also shown in Fig. 3 for comparison. As it is seen from Fig. 3, $\varepsilon''$ of PbSe QDs has a peak in the wavenumber range of 5000 cm$^{-1}$ – 10000 cm$^{-1}$ (depending on the diameter of the QDs) which does not exist in $\varepsilon''$ of the bulk. The imaginary part of the dielectric function represents absorption by the material. The dominant absorption mechanism in bulk lead chalcogenides (which are semiconducting materials) in this range of wavenumbers is the intrinsic absorption, i.e., the transition of electrons from the valence band to the conduction band due to absorption of photons with energy equal to or greater than the bandgap energy[154]. For this reason, $\varepsilon''$ of bulk lead chalcogenides abruptly increases at the bandgap energy (~2250 cm$^{-1}$ for PbSe) above which photons can generate electron-hole pairs. Unlike bulk lead chalcogenides, $\varepsilon''$ of the QDs is negligible for wavenumbers slightly larger than the bandgap causing a peak in $\varepsilon''$ at the bandgap energy. Quantum dots have negligible $\varepsilon''$ above the bandgap energy since their energy levels are discrete due to the quantum confinement effect. As such, photons with energy slightly higher than the bandgap are less likely to be absorbed as their energy needs to exactly match the difference between two energy levels in the valence and conduction bands. This causes $\varepsilon''$ of the QDs to peak at the bandgap energy of the QDs. The bandgap energy of the QDs is strongly dependent on their size and so is the peak of their $\varepsilon''$. The peak of $\varepsilon''$ of the QDs significantly shifts toward smaller wavelengths as the diameter of the QDs increases. It is seen from Fig. 3a that $\varepsilon'$ increases to a maximum value around the wavenumber of the peak of $\varepsilon''$ and then rapidly drops. This phenomenon is known as the anomalous dispersion[155]. Using the extracted dielectric functions, the spectral LDOS due to thermal radiation of periodic arrays of PbS, PbSe, and PbTe QDs is computed at an observation distance $d$ of 50 nm. The QDs in the arrays are separated by a distance of $4D$ in both $x$- and $y$-directions (i.e., the array pitch is $L = 5D$). Since the diameter of the QDs is much smaller than the wavelength, the observation distance and the interdot spacing, there is no need to discretize the QDs into smaller sub-volumes. The calculated LDOS spectra for PbS, PbSe, and PbTe QDs are shown in Figs. 4(a), 4(b), and 4(c), respectively. It is seen from Fig. 4 that the LDOS of the QD arrays has a peak whose spectral location strongly depends on the diameter of the QDs. The peak in the PbS spectrum is located at 5714 cm$^{-1}$ when $D = 6.8$ nm. As $D$ decreases, the spectral location of the peak shifts greatly toward higher frequencies such that the peak wavenumber reaches 10204 cm$^{-1}$ (a shift of 4490 cm$^{-1}$) when $D$ decreases by only 3.1 nm (i.e., when $D$ = 3.7 nm). Similar size-dependent spectra are observed for periodic arrays of PbSe and PbTe QDs. The peak location of the LDOS thermally emitted by the PbSe (PbTe) QD arrays shifts from 5952 cm$^{-1}$ (5682 cm$^{-1}$) for a diameter of 5.7 nm (6.7 nm) to 8265 cm$^{-1}$ (6969 cm$^{-1}$) for a diameter of 3.3 nm (4.9 nm). The peaks observed in the LDOS spectra can be explained by considering the fact that thermal radiation by the QDs is proportional to the imaginary part of their electric polarizability given by the Clausius–Mossotti relation as $\alpha'' = \dfrac{3\pi D^3 \varepsilon''}{2|\varepsilon+2|^2}$. It is seen that $\alpha''$ is directly proportional to the imaginary part of the dielectric function, $\varepsilon''$, which represents absorption by the QDs. As discussed before, $\varepsilon''$ peaks at the bandgap energy of the QDs (see Fig. 3). As such, $\alpha''$ and LDOS also exhibit a peak at the bandgap energy of the QDs. The bandgap energy of the QDs is strongly dependent on their diameter and so is the peak of thermal LDOS. The smaller the QDs, the larger the quantum confinement and the bandgap. As such, the peak of LDOS blue shifts significantly as QDs become smaller.

It should be noted that changing the pitch (or the filling factor) of the array mostly affects the magnitude of the LDOS rather than the spectral location of its peak. For example, the LDOS spectra for two arrays of 5.7-nm PbSe QDs with pitches of $5D$ ($f$ = 0.021) and $6D$ ($f$ = 0.015) are compared in Fig. 5. The magnitude of the LDOS for a pitch of $6D$ is about 30% less than that for a pitch of $5D$ throughout the considered wavenumber range. The reduction in the magnitude of the LDOS for the array with $L = 6D$ is due to the smaller density of QDs in this array. The wavenumber of the LDOS peak, however, remains unchanged with increasing the array pitch from $5D$ to $6D$. It should be noted that the coupling between neighboring QDs is negligible for both arrays with $L = 5D$ and $L = 6D$ due to the large interdot spacing. The coupling between neighboring QDs, which becomes significant for very dense arrays (where the interdot spacing $\ll D$), does not change the spectral location of the peak LDOS.

The effect of increasing the number of QD layers in the array on the magnitude and the spectrum of the LDOS is also studied. The LDOS thermally emitted by an array made of $N_L$ layers of PbSe QDs (see the inset of Fig. 6) is shown in Fig. 6 for $N_L$ ranging from 1 to 20. The QDs have a diameter of $D = 5.7$ nm, while the pitch size of the QD layers and the interlayer spacing are fixed at $5D$. The LDOS is calculated at an observation distance of $d = 50$ nm above the top layer. It can be seen from Fig. 6 that adding QD layers to the array increases the magnitude of the emitted LDOS. The contribution of additional layers to the emitted LDOS decreases with increasing $N_L$ such that increasing $N_L$ beyond 10 does not affect the LDOS anymore. The magnitude of the total (spectrally integrated over a wavenumber range of 5000 – 9000 cm$^{-1}$) LDOS for $N_L = 10$ is 33.1% larger than that for an array with $N_L = 1$. The spectrum of the LDOS and the wavenumber of the LDOS peak, however, remains unchanged as additional layers are added to the array.

It should be noted that the peak of near-field radiative heat transfer between two identical arrays of QDs can also be strongly tuned by changing the diameter of the QDs. The peak of thermal radiation by the emitting array is located at the same wavenumber as the peak of absorption by the receiving array (both are located around the bandgap energy of the QDs). As such, the near-field radiative heat transfer between the two arrays also peaks around the bandgap energy of the QDs. As an example, near-field radiative heat flux between two periodic arrays of PbSe QDs with a pitch size of $5D$ separated by a gap of $d = 50$ nm is computed using the T-DDA for three different QD diameters. The emitting array is at 1200 K, while the receiving array is kept at 300 K. The near-field radiative heat flux between the two arrays is shown in Fig. 7 for three QD diameters of 4.7 nm, 5.2 nm, and 5.7 nm. The peak of near-field radiative heat transfer shifts significantly from 6725 cm$^{-1}$ to 5925 cm$^{-1}$, when $D$ slightly increases only from 4.7 nm to 5.7 nm.

Quantum dots can be fabricated in large quantities using cost-effective solution-based methods. The solution-based fabrication techniques result in random, rather than periodic, arrays of QDs. To study how the magnitude and spectrum of thermal radiation by random arrays are different from the periodic ones, the thermal LDOS for 100 random monolayer arrays is modeled using the non-periodic T-DDA[156]. The PbSe QDs with a diameter of 5.7 nm are considered for this study. The number of QDs in the random arrays is selected such that the filling factor of these arrays is the same as that of a periodic array with $L = 6D$ (i.e., $f = 0.015$). Similar to the periodic array, a minimum interdot spacing of $5D$ is assumed between the QDs. The LDOS is calculated at a distance of $d = 50$ nm above the array. Initially, the LDOS is calculated for a $5L$ by $5L$ random array of QDs. Then, the array size is increased until no considerable change in the LDOS calculated at the distance of $d = 50$ nm is observed. An array size of $20L$ by $20L$ is found to be sufficient for modeling the LDOS at $d = 50$ nm. The spectral LDOS for the random arrays is compared with that for the periodic array in Fig. 8. As it is seen from this figure, the LDOS spectra and peak locations of the random arrays are similar to that for the periodic one. However, the magnitude of the LDOS fluctuates by 26% around its value for the periodic array. Figure 8 shows that the spectrum of near-field thermal radiation can be tuned using random arrays of QDs which can be fabricated in large quantities using cost-effective techniques.

## IV. CONCLUSIONS

The thermally emitted LDOS by lead chalcogenide QD arrays at a near-field distance from the array as well as near-field radiative heat transfer between two identical lead chalcogenide QD arrays were calculated for various QD sizes using the T-DDA. The size-dependent dielectric function of the QDs was extracted from the measured absorption spectra found in the literature using the Maxwell-Garnett EMT, the KK relation and an optimization technique. It was shown that the near-field LDOS and heat flux have a peak around the bandgap energy of the QDs which strongly varies with the size of the QDs due to the quantum confinement effect. This peak is due to the large imaginary part of the polarizability of the QDs at the bandgap energy where interband transitions is the dominant absorption mechanism. This study shows that the spectrum of near-field thermal radiation can be tuned by capitalizing on the size-dependent quantum confinement effect.

**Acknowledgments**

This work is supported by the National Science Foundation under Grant No. CBET-1804360.

**Data Availability**

The data that support the findings of this study are available from the corresponding author upon reasonable request.

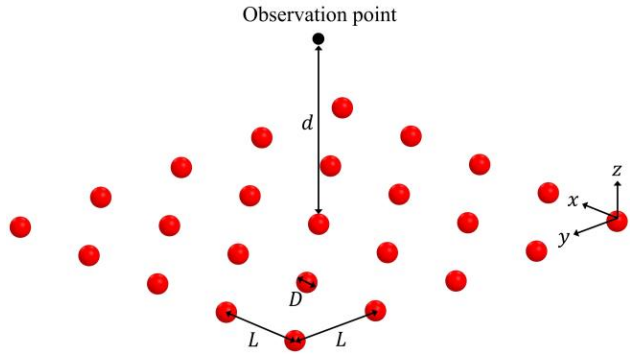

Figure 1 – A periodic array of QDs with diameter $D$ is thermally emitting into the free space. The array has a pitch $L$ along the $x$- and $y$-directions. The LDOS at an observation point located at a distance $d$ above the array and along the central axis of the QDs is of interest.

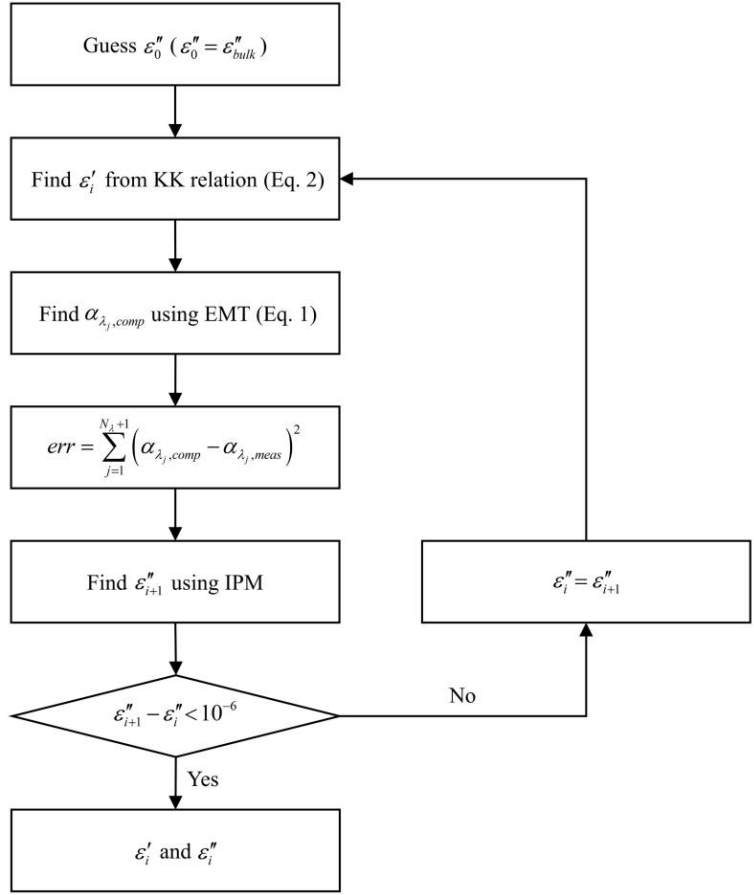

Figure 2 – A block diagram of the technique used for extracting the dielectric function of the QDs from their absorption coefficient. $\varepsilon'_m$ and $\varepsilon''_m$ ($m = i$, $i+1$) refer to the real and imaginary parts of the dielectric function of the QDs in iteration $m$, respectively.

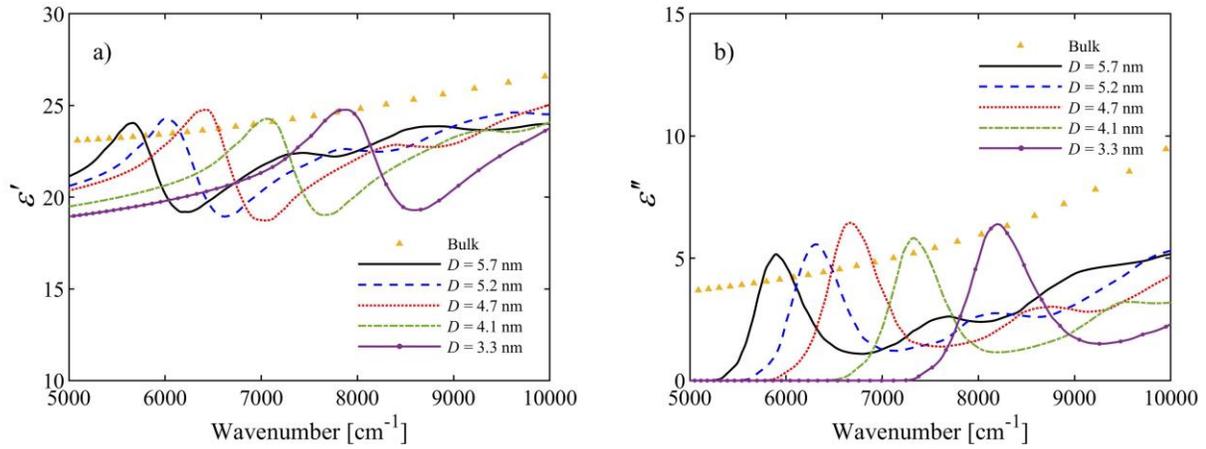

Figure 3 – The (a) real and the (b) imaginary parts of the size-dependent dielectric function of PbSe QDs in comparison with the dielectric function of a bulk of PbSe. The QDs have a diameter of $D$.

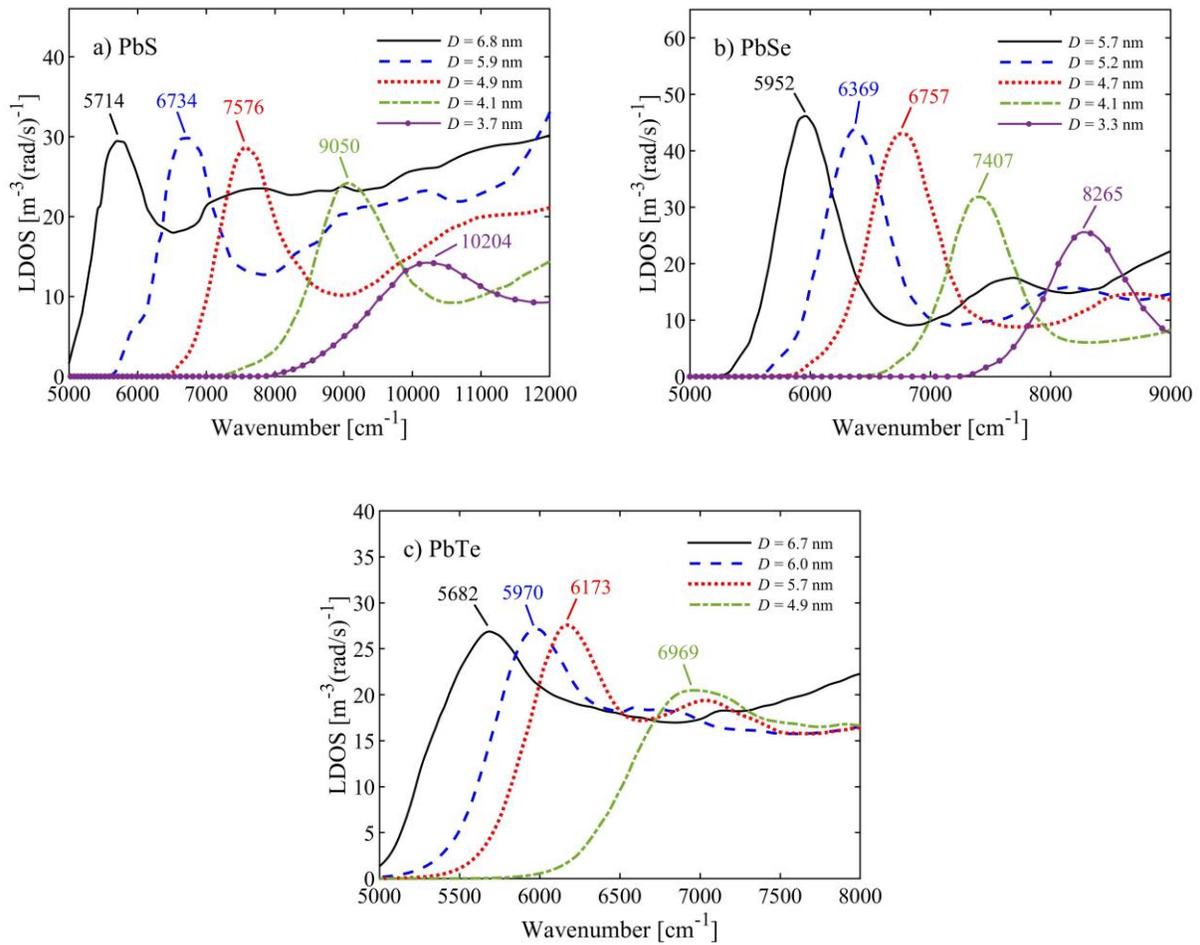

Figure 4 – Spectral LDOS thermally emitted by periodic arrays of (a) PbS, (b) PbSe, and (c) PbTe QDs with various diameters $D$ at an observation distance of $d = 50$ nm above the array along the central axis of the QDs. The array pitch is $5D$.

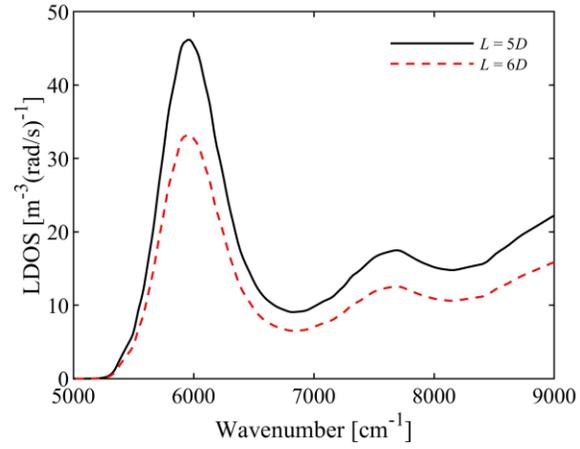

Figure 5 –The effect of array pitch on the thermally emitted LDOS. The spectral LDOS is calculated for periodic arrays of PbSe quantum dots with a diameter of $D = 5.7$ nm and pitches of $L = 5D$ and $L = 6D$ at an observation distance of $d = 50$ nm.

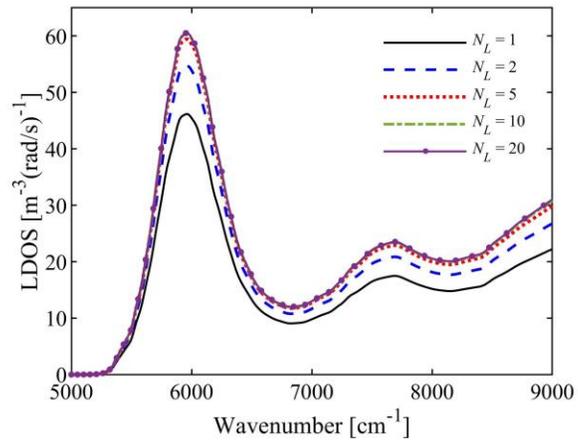

Figure 6 – The thermal LDOS emitted by $N_L$ periodic layers of PbSe QDs with a diameter of $D =$ 5.7 nm at an observation distance of $d =$ 50 nm above the top layer. The pitch size of the periodic layers and the interlayer spacing are $5D$.

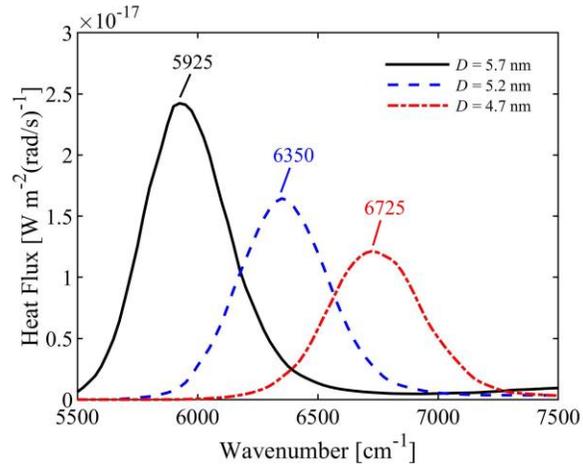

Figure 7 – Spectral heat flux between two periodic arrays of PbS QDs with various diameters *D* and an array pitch of 5*D* separated by a distance of 50 nm. The temperatures of the emitting and receiving arrays are 1200 K and 300 K, respectively.

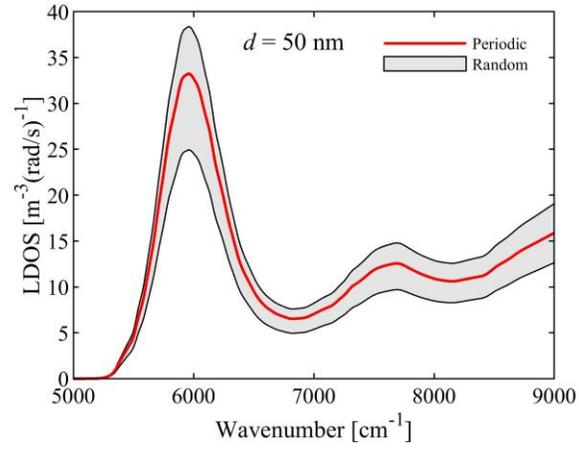

Figure 8 – Range of change in spectral LDOS thermally emitted by a randomly distributed array of PbSe quantum dots with a diameter of $D$ = 5.7 nm and a filling factor of $f$ = 0.015 relative to that emitted by a periodic array of the same $D$ and $f$.